# Multi-channel secure communication framework for wireless IoT (MCSC-WoT): enhancing security in Internet of Things

Prokash Barman[1] · Ratul Chowdhury[2] · Banani Saha[1]

## Abstract
In modern smart systems, the convergence of the Internet of Things (IoT) and Wireless of Things (WoT) have been revolutionized by offering a broad level of wireless connectivity and communication among various devices. Hitherto, this greater interconnectivity poses important security problems, including the question of how to securely interconnect different networks, preserve secure communication channels, and maintain data integrity. However, the traditional cryptographic method and frequency hopping technique, although they provide some protection, are not sufficient to defend against Man-In-The-Middle, jamming, and replay attacks. In addition, synchronization issues in multi-channel communication systems result in increased latency and energy consumption, which make them unsuitable for resource-constrained IoT and WoT devices. This work presents the Multi-Channel Secure Communication (MCSC) framework, which integrates advanced cryptographic protocols with dynamic channel-hopping strategies to enhance security with reduced synchronization overhead. The MCSC framework maximizes the critical performance metrics, such as packet delivery ratio, latency, throughput, and energy efficiency, and fulfills the specific requirements of the IoT and WoT networks. A comprehensive comparison of MCSC with well-established methods, including Frequency Hop Spread Spectrum, single channel Advanced Encryption Standard, and various Elliptic Curve Cryptography-based schemes, indicates that MCSC has lower error rates and is more resilient to a wider range of cyber attacks. The efficiency of the proposed solution to secure IoT and WoT networks without compromising the operational performance is validated under various interference conditions.

Ratul Chowdhury and Banani Saha have contributed equally to this work.

✉ Prokash Barman
pbcse_rs@caluniv.ac.in

Ratul Chowdhury
ratul.chowdhury@nsec.ac.in

Banani Saha
bscomp@caluniv.ac.in

1 Department of Computer Science and Engineering, University of Calcutta, Saltlake City, Kolkata, West Bengal 700091, India

2 Department of Computer Science and Engineering (AIML), Netaji Subhash Engineering College,  Garia, Techno City, Kolkata, West Bengal  700152, India

## 1 Introduction

The area of the Internet of Things and Wireless of Things have progressed rapidly and reshaped how things communicate in many applications such as smart homes, healthcare, industrial automation, and environmental monitoring. IoT systems enable real-time data exchange, remote control, and automation by allowing interconnected devices to communicate with each other wirelessly. But, greater connectivity means more vulnerabilities in IoT/WoT devices, which are typically resource-constrained and operate in an open wireless environment. However, the constraints make them vulnerable to several attacks such as eavesdropping, jamming, and interference, which can degrade the integrity of data as well as the network availability. Securing communication between devices is one of the primary challenges of IoT/WoT systems, and it needs to be done without introducing too significant a burden in terms of computation or energy. Traditional methods like RSA [24] and Elliptic Curve Cryptography [16] are known for offering good security guarantees but come at an expensive in terms of processing power and energy consumption, making them inadvisable for low-power devices of IoT. With that, as IoT networks scale up, there is more and more demand for lightweight security solutions, finding the right balance between strong encryption and energy efficiency.

The Multi-Channel Secure Communication framework, a comprehensive security solution that protects wireless communication in WoT environments, is proposed in this paper to address these challenges. The MCSC framework combines AES [11], a proven efficient and secure encryption scheme, with a dynamic, multi-channel hopping mechanism to protect against the most common wireless attacks, e.g. jamming and eavesdropping. In this context, the use of AES is particularly relevant since it achieves a favorable balance between security strength and computational efficiency and is suitable for energy-constrained IoT devices. Moreover, a dynamic multi-channel hopping mechanism is introduced, which acts as another randomized aspect in communication, thus making it more difficult to disrupt or intercept communications. This work is motivated by the fact that existing work of securing IoT communication either jeopardizes security for conserving energy or imposes prohibitively high computational cost for devices. Additionally, cryptographic techniques come with energy and computational load overhead, and mechanisms such as frequency hopping or channel hopping are themselves complex and require costly synchronization in constrained environments. To close these gaps, the MCSC framework leverages a relatively lightweight mechanism for synchronization, designed to minimize the energy cost while providing a resilient defense against passive and active attacks. MCSC is a dynamic channel-hopping mechanism that guarantees communication over multiple frequency channels in a random pattern. Because it is very hard for an attacker to predict or trace the channel sequence, this process greatly lowers the chance of an attack successfully jamming or intercepting the channel. Unlike traditional channel hopping systems that need complex synchronization protocol, our MCSC comes with a lightweight synchronization mechanism and reduces the computational overhead and energy consumption of channel switching. In addition to the MCSC channel-hopping mechanism, encryption is achieved by using AES-based encryption for data transmission between IoT devices. Certainly, AES is widely used due to its security, speed, and efficiency balance. Other encryption schemes, such as ECC, are more complex and not well suited to IoT environments, while AES offers a relatively simple and energy-efficient means of encryption for use in low-power IoT environments. This paper compares AES and other cryptographic methods in detail and explains why AES was chosen for this framework. In particular, AES offers good security with very low computational and energy requirements and is well suited to the constrained resources within an IoT device.

The proposed framework incorporates the process of Packetization (structuring) of the sent data. This packet structure used in the MCSC framework minimizes the overhead and guarantees the secure transmission. We evaluate the performance metrics of the MCSC framework, including latency, packet delivery ratio, and energy consumption in this work. These metrics are essential in evaluating the feasibility of using the framework for real world IoT applications. In IoT, latency is the time delay in communication and is especially critical in time-sensitive IoT applications such as healthcare monitoring and industrial automation. PDR measures the communication reliability over the percentage of packets successfully delivered. The dynamic channel-hopping process directly impacts both metrics, as frequent channel switching can lead to delay or increase the chance of packet loss. Nevertheless, the lightweight synchronization mechanism devised in MCSC helps to minimize these performance trade-offs. This paper proposes a novel approach, which will incorporate AES encryption with a hopping mechanism that is both dynamic and multi–channel. The proposed MCSC framework mitigates some key security challenges and issues in the IoT environment, such as the requirement for low latency and energy-efficient communication protocols.

The contributions of this work are as follows:

1. A lightweight synchronization mechanism that enables efficient and secure channel switching that is suitable for resource constrained environments.
2. Comprehensive analysis of AES encryption as an optimal choice for IoT devices vs. other cryptographic schemes.
3. Improved technique that offers a secure, randomized multi-channel hopping process to robustly fend off against jamming and eavesdropping.
4. It provides an in-depth performance evaluation that proves the effectiveness of the framework concerning latency, PDR, and energy consumption.

The structure of this paper is as follows: Sect. 2 provides a comprehensive review of the literature, discussing existing methods and their limitations. In Sect. 3 MCSC framework, its design, algorithms, and key components are elaborated. Section 4 describes the hardware implementation of the framework. Results and performance analysis of the framework based on key metrics such as latency, energy consumption, and packet delivery ratio are presented in Sect. 5. Practical application of the proposed framework depicted in Sect. 6. Implications of the findings with limitations of the current system, and suggestion for future research directions are discussed in Sect. 7. Finally, Sect. 8 concludes the paper with a summary of the contributions and potential impact of the proposed framework.

## 2 Related work

In recent years, the flood of IoT networks has turned the security of these networks into a focal point as these devices are extremely vulnerable to most security threats. This review encompasses various contributions to IoT security, especially in cryptographic mechanisms, frequency-hopping/channel-hopping techniques, and synchronisation techniques. The proposed Multi multi-channel secure Communication framework is based on a comparison table that indicates the strengths and weaknesses of existing studies.

### 2.1 Cryptographic solutions for IoT

The cryptographic security in IoT networks is critical for the end to end (E2E) communications. Zhang et al. [31] proposed a lightweight AES encryption scheme for IoT devices, that offers both security and energy efficiency. This approach, however, does not consider jamming attacks, which are severe threats in wireless scenarios. In Mahmood et al. [21] studied an AES and ECC hybrid encryption model that improved security whilst keeping computational overhead minimal. Though the hybrid solution has its advantages, it's still resource-hungry and less appropriate for ultra-low power devices. Probabilistic encryption as looked at by Benaloh [7], also provides some early work in the foundation of lightweight security solutions in comparison to current solutions (such as AES). Identity-based encryption [9], reduces the computational requirements associated with public key infrastructures, but lacks sufficient resource efficiency for use in IoT use cases. Aiming at providing general guidelines for security in WSN, Wang et al. [28] made the survey well suited for the development of cryptographic strategies for IoT. Unlike this, Ren et al. [23] tackled multiuser broadcast authentication in the context of wireless sensor networks, appropriate for safe communication in IoT. Moosavi et al. [22] proposes a highly similar secure end to end scheme for mobile healthcare IoT, which is comparable to goals of the MCSC framework. In Ghosh et al. [14], energy-efficient security communication schemes are highlighted, which is the need for security and energy efficiency for IoT devices balancing. While most of these cryptographic solutions are a vital foundation for securing low power devices, they do not typically consider the threat of jamming attacks nor power or resource constraints.

### 2.2 Frequency-hopping and channel-hopping mechanisms

To overcome the problems related to the use of fixed channels, frequency hopping spread spectrum and channel hopping techniques have been introduced. In particular, Wu et al. [29] presented a lightweight symmetric encryption scheme that leverages FHSS in order to protect against eavesdropping. While FHSS allows for a more complex synchronisation, thus introducing increased latency. There have been proposals to address jamming attacks on industrial IoT through the use of FHSS [26], but the synchronisation problem degrades performance, particularly in resource-constrained environments. The need for communications security in wireless networks, especially in dynamic, multi-channel environments such as IoT, is stressed in Bhargava and Poor [8]. In 2020, Liu et al. [20] introduced an energy-efficient channel-hopping mechanism that supports low-power devices within ISM. Yet, synchronisation overhead remains a problem. Ahmed et al. [1] applied cognitive radio for dynamic channel selection in the presence of varying interference levels such that channel utilisation should increase at the expense, however, of additional computational complexity for IoT devices. In earlier work, the security of distributed sensor networks (precursor to IoT solutions) was done by Carman et al. [10]. The channel-hopping methods introduced in Ahmed et al. [1] by using the cognitive radio-based approach show how dynamic channel selection can increase network resilience but at the cost of computation.

### 2.3 Synchronization in multi-channel communication

Synchronisation is a critical challenge in multi-channel communication systems. A clock synchronization algo-rithm proposed by Liu et al. [19] achieved improved precision in multi-channel communications but with higher latency because of channel switching. Sun et al. [27] proposed a lightweight synchronisation protocol using periodic synchronisation signals to reduce computational overhead at the cost of energy due to the high-frequency synchronisation. In Jiang et al. [15], we developed a scheduling-free communication system where devices pick channels based on pre-determined sequences to remove the explicit synchronisation. On the other hand, this approach is susceptible to desynchronization caused by packet loss, rendering the communication redundant. A useful idea that comes from Lazos and Poovendran [18] is a robust localization protocol to wireless sensor networks and provides resources on dealing with synchronisation issues in low-power systems. Carman et al. [10] pointed out that synchronisation overhead is still an interesting issue in IoT systems, particularly for multi-channel protocols that care about energy efficiency. A number of solutions in the literature allow for various trade-offs among energy consumption, latency, and synchronisation precision but show no clear optimal solution for low-power IoT devices.

## 2.4 Authentication approaches

The recent development of privacy-enhancing and authentication methods for 5 G-enabled vehicle fog computing effectively tackles major safety issues. The authors of Almazroi et al. [6] describe a bilinear pairing system that uses anonymous authentication but preserves user privacy through reduced computational costs. Al-Mekhlafi et al. [3] introduced CLA-FC5G as a certificateless authentication technique that uses fog computing together with device-to-device communication for providing scalable and efficient solutions to vehicle networks. The authors Almazroi et al. [5] implemented Chebyshev polynomials within a lightweight cryptographic solution to minimize DDoS attacks targeting 5 G-based automotive networks. Al-Mekhlafi et al. [2] achieved another milestone by integrating elliptic curve cryptography with blockchain for implementing certificateless authentication in vehicular communications. The oblivious transfer-based protocol developed by Al-Mekhlafi et al. [4] achieved higher privacy through vehicle request hiding with lower expenses while providing better security. The new technological solutions boost 5G-enabled vehicular network security as well as scalability and operational efficiency.

## 2.5 Limitations of existing approaches

Different IoT security approaches reveal strengths and weaknesses in this literature review. In general, most of the cryptographic solutions like AES give high security but alleviate against jamming attacks. Eavesdropping protection with frequency-hopping techniques comes at the expense of synchronisation complexity and, thus, additional latency and energy consumption. Xu et al. [30] also analyzed jamming sensor networks, discussing attack strategies and defences, and this applies to evaluating the vulnerabilities in current solutions. On the other hand, as pointed out by Zhang et al. [32], synchronisation and energy efficiency issues also affect the security of smart city IoT systems. Other additional insights to energy-efficient communication schemes for low-power devices were provided by Ghosh et al. [14], however, there still existed synchronisation overhead inherent in these systems. Broadcast authentication methods [23] are essential for secure multi-channel communication, but energy consumption remains a bottleneck in many systems.

## 2.6 Research gap and motivation for MCSC

The Table 1 summarizes the existing solutions reviewed in this section, outlining their techniques, strengths, and weaknesses.

With all the advancements in IoT security, existing solutions have to trade off between security, energy efficiency, and performance. To address these challenges, we propose the Multi-Channel Secure Communication (MCSC) framework, which integrates a lightweight synchronization mechanism that reduces energy consumption and a randomized multi-channel communication scheme that provides robust protection against jamming and eavesdropping. MCSC works effectively with low overhead, thus being suitable for low-power IoT devices.

The MCSC framework is grounded on past research but responds to the important gaps observed in the literature. In the following sections, we present the design, implementation, and performance evaluation of the MCSC framework, which we show to be effective in improving security for resource-constrained IoT environments.

## 3 Proposed MCSC framework

The objective of the Multi-Channel Secure Communication framework is to provide a secure and efficient communi-cation solution within WoT-related environments. AES encryption, multi-channel communication, a packetization module, and a lightweight synchronisation mechanism are incorporated here. The detailed descriptions of each module are elaborated in the following subsections.

Table 1 Principal methods, their strengths, and weaknesses, from the reviewed literature

| Paper | Authors | Approach | Strengths | Limitations |
| --- | --- | --- | --- | --- |
| [29] | Wu et al. | Lightweight symmetric encryption + FHSS | Protection against eavesdropping | Complex synchronization increases latency |
| [31] | Zhang et al. | Lightweight AES | Strong security, energy-efficient | Vulnerable to jamming attacks |
| [1] | Ahmed et al. | Cognitive radio channel hopping | Dynamic channel selection | Increased computational burden |
| [20] | Liu et al. | Channel-hopping mechanism | Improved energy efficiency | High synchronization overhead |
| [19] | Li et al. | Clock synchronization algorithm | Improved synchronization precision | Increased latency during channel switching |
| [21] | Mahmood et al. | Hybrid AES and ECC | Enhanced security, reduced computational overhead | Requires significant resources |
| [14] | Ghosh et al. | Energy-efficient communication schemes | Balances security and energy | High synchronization cost |
| [15] | Jiang et al. | Synchronization-free system | Eliminates explicit synchronization | Vulnerable to desynchronization |
| [26] | Smith et al. | FHSS-based security framework | Mitigates jamming attacks | Synchronization challenges lead to performance issues |
| [27] | Sun et al. | Lightweight synchronization | Reduced computational overhead | Energy consumption during high-frequency sync |
| [22] | Moosavi et al. | E2E security for healthcare IoT | Secure healthcare communications | Energy consumption |
| [23] | Ren et al. | Multi-user broadcast authentication | Secure communication for low-power devices | Energy overhead in authentication |
| [18] | Lazos and Poovendran | Robust localization for sensor networks | Improves synchronization and localization | Performance drops in high-power environments |
| [30] | Xu et al. | Jamming defence | Attack strategy defence | Synchronization challenges |
| [9] | Boneh and Franklin | Identity-based encryption | Public key infrastructure improvements | Resource-heavy for IoT |
| [7] | Benaloh | Probabilistic encryption | Early innovations in encryption | Does not address jamming attacks |
| [17] | Koblitz | Elliptic Curve Cryptosystems | High security with smaller key sizes | More complex than symmetric encryption |
| [25] | Rivest et al. | RSA Cryptosystem | Strong security for public key infrastructure | Computationally intensive, unsuitable for low-power |
| [12] | Diffie and Hellman | Key Exchange Protocol | Secure key exchange between parties | Vulnerable to man-in-the-middle attacks |

The MCSC framework described in Fig. 1 consists of four key modules.

1. AES Encryption module: This module manages data confidentiality in connection with transmission.
2. Multi-channel communication module: Presents a feature for randomized channel change to raise security levels and lessen the threat of eavesdropping and jamming.
3. Synchronization module: This module synchronize both parties to keep aligned in their communication, even during the exchange of different communication channels.
4. Packetization module: This module deployed in the process of gathering data, organizing it in structured packets for delivery, and fuses control alongside encrypted components.

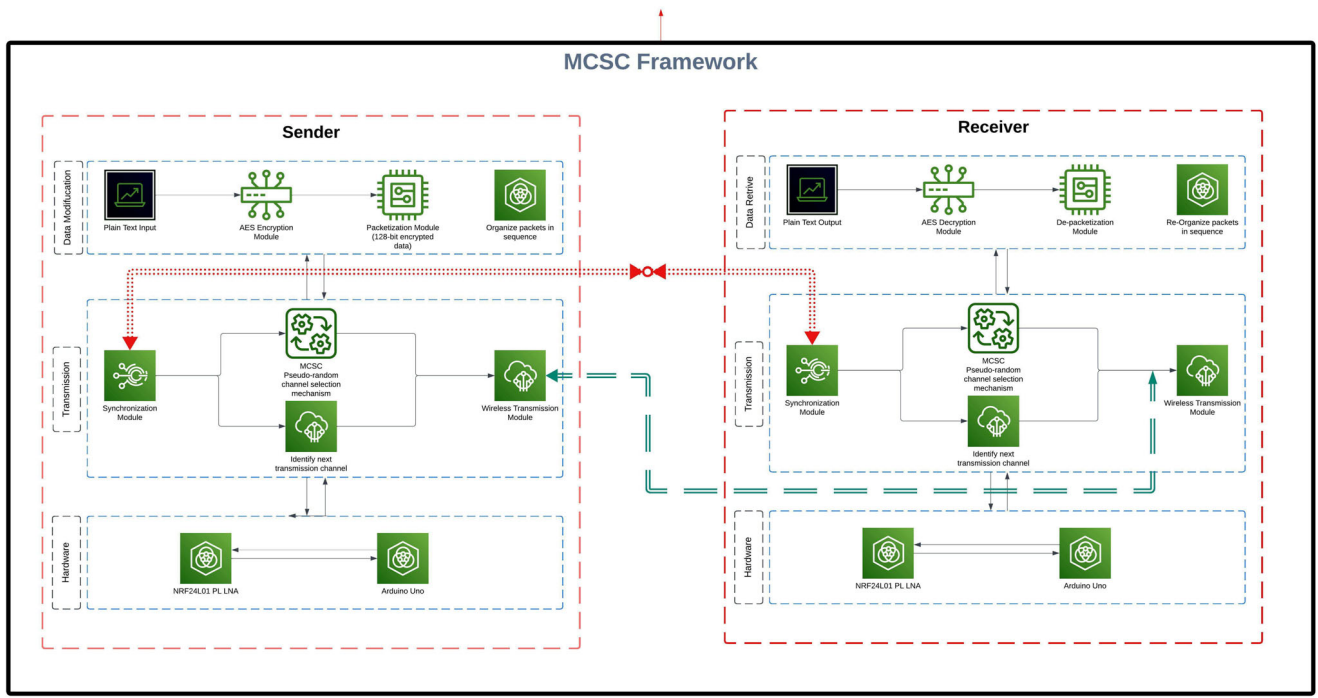

**Fig. 1** MCSC framework comprising different components

## 3.1 AES encryption module

The MCSC framework provides an AES Encryption module that depends on the Advanced Encryption Standard for securing IoT and WoT device interconnections. Symmetric AES encryption demonstrates excellent security features together with fast performance and efficient operation which makes this algorithm best suited for devices having restricted power capabilities and minimal processing resources. AES enables the secure encryption of data into protected formats that provide confidentiality through its encryption and maintains fewer computational costs than advanced encryption techniques, including Elliptic Curve Cryptography . AES encryption becomes stronger by using dynamic multi-channel hopping because this combination secures transmitted data and complicates hacker access to communication channels, thus protecting against eavesdropping attacks and jamming and man-in-the-middle interference attempts. The framework achieves effective IoT and WoT communication protection through AES encryption because it integrates with dynamic channel-hopping functionality without energy or performance degradation.

## 3.2 Multi-channel communication

After encryption, this module facilitates communication across multiple dynamically selected channels, which in turn reduces the chance of eavesdropping and jamming attacks. This algorithm randomly switches between channels (Channel hopping) in a secure and unpredictable manner using a Pseudorandom Number Generator (PRNG). The algorithm leverages a shared seed and PRNG to generate a secure, synchronized hopping sequence for all nodes. The detailed descriptions of the intermediate two steps channel hopping and PRNG is defined the following sections.

### 3.2.1 Channel hopping

The mechanism used by the MCSC framework for channel hopping allows the sender and receiver to switch between channels in a coordinated manner based on a common pseudo-random sequence. The *Multi-Channel Hopping* mechanism is described in the following Algorithm 1.

**Algorithm 1** Multi-Channel Hopping with PRNG

1: **Input:**
2: 　　$C_{list}$: List of available channels.
3: 　　$N$: Total number of channels.
4: 　　$T_{sync}$: Synchronization interval.
5: 　　$Seed$: Shared seed for the PRNG.
6: 　　$P_{generator}$: Pseudorandom Number Generator.
7: **Output:** Synchronized, secure channel hopping sequence.
8: **procedure** MULTI-CHANNEL HOPPING()
9: 　　**Initialization**
10: 　　　Initialize PRNG with $Seed$.
11: 　　　Generate an initial sequence $S_{initial} = P_{generator}(Seed, N)$.
12: 　　　Assign $T_{current}$ based on system clock.
13: 　　**Channel Selection**
14: 　　　For each $T_{current}$:
15: 　　　$Index = P_{generator}(Seed + T_{current}, N)$.
16: 　　　$C_{current} = C_{list}[Index]$.
17: 　　　Switch to channel $C_{current}$.
18: 　　**Data Transmission**
19: 　　　Transmit or receive data on $C_{current}$.
20: 　　**Periodic Synchronization**
21: 　　　At each $T_{sync}$, synchronize PRNG and update $Seed$ if necessary.
22: 　　**Resynchronization**
23: 　　　IF (Synchronization signal lost)
24: 　　　　Select random channel $C_{resync}$ from $C_{list}$.
25: 　　　　Wait for next sync signal on $C_{resync}$.
26: 　　　ENDIF
27: **end procedure**

The Algorithm 1 depicts the multi-channel hopping with the PNRG mechanism. Whereas the Table 2 describes different parameters used in the Algorithm 1. The Algorithm 1 takes a list of available communication channels, synchronization interval, and seed for Pseudorandom Number Generator as input and provides a secure channel hopping sequence in a synchronized manner. The algorithm passes through different phases like initialization, channel selection, data transmission, and synchronization. In the following sections, different functions of the Algorithm 1 are elaborated.

**Table 2** Parameters used in multi-channel hopping algorithm

| Parameter | Description |
| --- | --- |
| $C_{list}$ | List of available communication channels in the network. Example: $[1, 2, 3, ..., N]$ |
| $N$ | Total number of available channels in $C_{list}$ |
| $T_{sync}$ | Synchronization interval. Defines how often synchronization signals are sent by the master node |
| $Seed$ | Initial seed shared among all nodes for the pseudorandom number generator (PRNG) |
| $P_{generator}$ | Pseudorandom Number Generator function used to generate the hopping sequence |
| $T_{current}$ | Current time based on the system clock, used to compute the next channel index |
| $S_{initial}$ | Initial sequence of channel indices generated by the PRNG during initialization |
| $Index$ | Index of the current channel in $C_{list}$, calculated using the PRNG and the current time |
| $C_{current}$ | Current channel selected for communication, derived from $C_{list}$ using the $Index$ |
| $C_{resync}$ | Randomly selected channel used for resynchronization if synchronization is lost |
| $T_{offset}$ | Time offset between the received synchronization signal and the local clock of the node |
| $Tmax_{offset}$ | Maximum allowable time offset for synchronization to be considered valid |

### 3.2.2 Pseudorandom number generator

This module is the second component of the MCSC framework. We applied the pseudorandom number generator with a shared secret key K that is known by devices via a secure exchange. The next communication channel in sequence is decided by a number generated by the PRNG. Receiving devices should be able to agree on the order of events; however, attackers should not be able to guess the sequence.

The PRNG can be represented as:

$C_i = PRNG(S) mod |n|$

where

- $C_i$ is the selected channel for the i-th packet.
- S is the shared seed for synchronization between sender and receiver.
- n is the number of available channels.

In the equation above $C_i$ specifies the selected random channel within the available channels (n) of the ISM band. Here, $C_i$ is retrieved for the next $i - th$ packet generated by the packetization module of MCSC framework. The equation uses a $PNRG()$ function fed with $S$ shared (seed) among the receiver and sender. The Algorithm 1 uses this $C_i$ as $C_current$ for channel selection.

The pseudorandom number generator ensures that the channel-hopping sequence is unpredictable, thus preventing eavesdropping or jamming attacks. A shared seed guarantees synchronization across all nodes. Regular synchronization, on the other hand, ensures that the hopping sequence remains consistent among all participating nodes. Nodes that lose synchronization can still rejoin the network by hopping to a random channel and waiting for the next synchronization signal. MCSC frameworks correspond to the problem of channel selection in multi-channel secure communications, which consists of choosing the next communication channel at random. The idea is to constantly hopping between multiple channels in a pseudo-random way to try and avoid jamming and eaves dropping. In this section we elaborate on the mathematical formulation of the channel selection mechanism.

Figure 2 shows the complete channel diagram of ISM Band (left) and nRF24L01 modules receiving channels (right). Overall 125 transmission channels of 1 MHz each are available between 2400 to 2525 MHz in the ISM band. The nRF24L01 has 6 available receiving pipes. Channel randomization at the sender end is made within the 125 available channels. In the receiver end, 6 receiving pipes are randomized

The equation for selecting the channel H(i) at time slot i based on the pseudorandom output is as follows:

$H(i) = (P(i) mod |C|) + 1$

where

- $P(i)$ is the pseudorandom number for the current time slot.
- C is the total number of available communication channels.
- mod C ensures that the result is within the valid range of channels, i.e., between 0 and $C - 1$.
- Adding 1 ensures that the channel numbers start from 1 instead of 0.

The equation presents $H(i)$ as pseudo-randomly chosen channel for transmission in $i$th time slot. The $H(i)$ is derived with pseudorandom number-$P(i)$ for the current time slot and the available communication channel-C. Adding 1 with the equation ensures that the channel number shall be started with 1 instead of 0. This equation guarantees that the selected channel H(i) is pseudo-. randomly chosen from the set of C available channels.

## 3.3 Synchronization module

For the sender and receiver to be synchronized on the same channel for a given time slot i, both must start with the same shared secret key K and be synchronized in terms of real-time (time slots). Communication failures can occur based on any synchronization mismatch. The synchronization module keeps the communication framework secured from attacks in the following ways:

- *Time-Varying Seed* The seed K is updated periodically with a cryptographic mechanism to boost the security and protect the secrecy of communication channels, thus, it happens that even an attacker if it gets to know the channel pattern for a short time, after the specified time the channels changed dynamically.
- *Frequency Hop Spread Spectrum (FHSS)* To prevent jamming attacks, frequency hopping among particular channels is combined with this channel selection technique, resulting in an unpredictable communication pattern for adversaries.

### 3.3.1 Synchronization mechanism

To synchronize the sending and receiving device clocks, the sending device keeps sending a synchronization packet regularly. The purpose is to reduce the time difference between different devices to make channel switching smooth. In MCSC frameworks, the Synchronization function guarantees that communicating devices (sender and receiver) are synchronized with time to select the same channel at the same moment. The system needs to operate

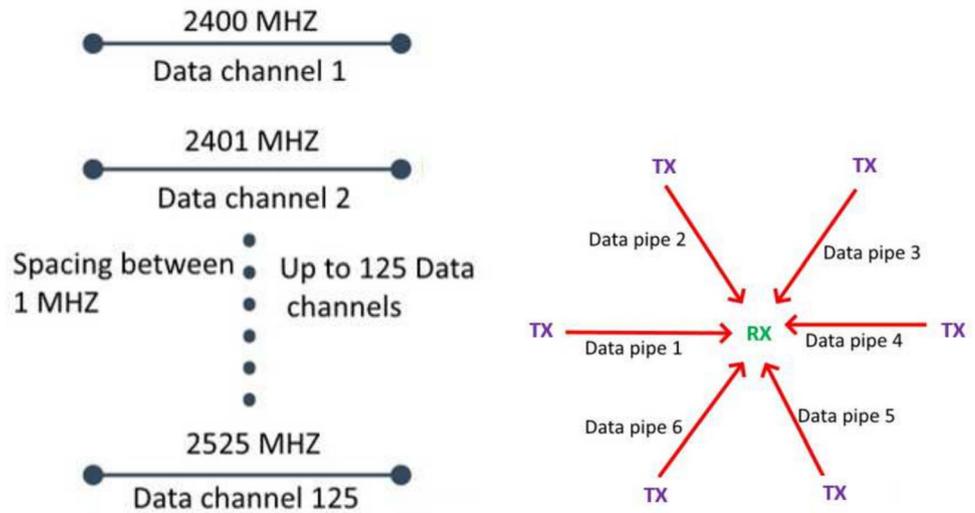

Fig. 2 Channel diagram in ISM Band and nRF24L01 (receiver)

correctly, as message breakdown would occur with de-synchronization. The main Purposes of this module:

1. All senders and receivers on the same communication channel.
2. Synchronization prevents packet loss due to the misalignment in time slots. Moreover, frequent resynchronization caused additional energy and computation costs, making it also relevant for low power and resource-constrained IoT devices to ensure efficiency to misalignment with time slots.
3. It also ensures efficiency in low-power and resource-constrained IoT devices, as frequent resynchronization incurs extra energy and computational costs.

### 3.3.2 Synchronization process

The internal clock of both the sender and receiver need to be synchronized for proper transmission in any time slot. The process of synchronization follow the Algorithm 2 and depends on four equations as follows:

1. *Clock Synchronization Mechanism* The Clocks of devices need to be synchronized. Its local time offset with respect to the network time is calculated by an equation. It is necessary to choose the same channel at the same time. The time offset $t_{offsett}$ between two devices is calculated as:

$$t_{offset} = |t_{receiver} - t_{sender}|$$

where

- $t_{receiver}$ is the time at the receiver's internal clock.
- $t_{sender}$ is the time at the sender's internal clock.

The requirement of clock synchronization arises if the value of $t_{offset}$ cross the predefined threshold limit. $t_{offset}$ is the difference between internal clock time at receiver ($t_{receiver}$) and sender($t_{sender}$).

2. *Maximum Clock Drift* The clocks on two devices can drift off from one another over time. The time difference that the system takes before resynchronization becomes necessary is defined as the maximum allowable clock drift Δt.

$$\Delta t = T_{sync} X drift_{rate}$$

where

- $t_{sync}$ is the time period after which resynchronization is performed.
- $drift_{rate}$ is the rate at which a clock drifts from the true time.

Predefined synchronization time interval ($t_{sync}$) and rate of time difference ($drift_{rate}$) between communicating end are fed into the equation to determine maximum allowable clock drift Δt.

3. *Synchronization Time Slot* For each communication time slot i, devices need to ensure they are synchronized within the acceptable drift range Δt. The equation to check synchronization for time the slot *i* is:

$$t_{receiver}(i) - t_{sender}(i) \leq \Delta t$$

The devices communicate on the same channel if the clock difference $t_{receiver} - t_{sender}$ is within Δt.

4. *Resynchronization* A resynchronization is required if the *clock drift* exceeds Δt. Synchronization time signals are emitted by the devices and each one accordingly adjusts its local clock. The new clock time ($t_{new}$) is defined as:

$$t_{new} = t_{old} + \frac{t_{receiver} - t_{sender}}{2}$$

where

- $t_{new}$ is the adjusted time after synchronization.
- $t_{old}$ is the time before synchronization.
- $t_{receiver}$ and $t_{sender}$ are the timestamps from the receiver and sender, respectively.

The synchronization process for the proposed framework is shown in Algorithm 2. The algorithm subdivided into 5 distinct operations, namely 1. Initialization, 2. Connection Establishment, 3. Sync Signal, 4. Channel Selection and Hopping and finaly 5. Resynchronization phases.

within an acceptable time offset $T_{max\_offset}$, the clock synchronization of all participating nodes is done after $T_{sync}$ interval. The resynchronization process in the algorithm prevents long-term desynchronization by allowing nodes to rejoin the network after missing signals.

The synchronization algorithm ensures that nodes in the MCSC system can hop between channels in a time-synchronized manner, minimizing the impact of time drift and reducing the risk of desynchronization.

### 3.4 Packetization module

This module breaks down data into consistent packets and fabricates them for transmission over the selected channels. Every packet consists of important control information

**Algorithm 2** Synchronization Algorithm

1: **Input:** $T_{sync}$, N, $S_{initial}$, $C_{list}$, $T_{max_{offset}}$, $R_{offset}$
2: **Output:** Synchronization among nodes
3: **procedure** SYNCHRONIZATION()
4:    **Initialization**
5:        Initialize $T_{current}$
6:        Set synchronization interval $T_{sync}$
7:        Determine set of available channels $C_{list}$.
8:    **Connection Establishment**
9:        Set T=$T_{sync}$, at master node.
10:       Broadcasts $S_{initial}$.
11:   **Process Sync Signal**
12:       Receive $S_{initial}$.
13:       Check $T_{offset}$.
14:       If $|T_{offset}| \leq T_{max\_offset}$ then go to Channel Switching.
15:       Else If $|T_{offset}| > T_{max\_offset}$, Sync local node's clock with the master node's clock.
16:   **Channel Selection and Hopping:**
17:       Select random channel, where $C_{current} = C_{list}[T_{current} \mod N]$.
18:       Switch to the next random channel in $C_{list}$ on every $T_{sync}$.
19:   **Resynchronize** after each interval $T_{sync}$, to adjust time drift.
20: **end procedure**

Parameters:

- T: Total time slot duration.
- i: Current time slot or communication round.
- N: Number of nodes in the network (in case of multi-node communication).
- $S_{initial}$: Broadcasts initial synchronization signal.
- $T_{sync}$: Synchronization interval or synchronization period.
- $t_{clock}$: Time in the sender/receiver's internal clock.
- Δt: Maximum clock drift or timing offset between devices.
- $t_{offset}$: Time difference between sender and receiver clocks.

Algorithm 2 is used by the master node to align the clocks of all participating nodes. To ensure that all nodes operate

along with an AES-encrypted payload. The Packet Structure contains the following fields:

1. *Next Channel Number (8 bits)* The destination through which the next packet will be sent.
2. *Node Address (16 bits)* As a tool for source identification, the address of the sending node serves a purpose.
3. *AES Encrypted Plain Text (88 bits)* Data in actual form gets encrypted with the help of AES.
4. *Packet Sequence Number (16 bits)* The function of the sequence number is to oversee the order of packets.

Figure 3 depicts the packet information as implemented in the proposed model.

The complete packetization process is defined in Algorithm 3. The module generates the final form of packets before transmission.

**Algorithm 3** Packetization Algorithm

```
 1: Input: Data_plain, N_c, Addr_node, K, S_pkt
 2: Output: Packet_enc
 3: procedure PACKETIZATION()
 4:     Initialization
 5:         Get Data_plain
 6:         Call SYNCHRONIZATION
 7:         Get N_c
 8:         Get Addr_node
 9:         Get AES encryption key K.
10:         Initialize S_pkt.
11:     AES Encryption
12:         Data_enc = AES-Encrypt(K, Data_plain)
13:     Generate Final Packet
14:         Packet_enc = |N_c|Addr_node|Data_enc|S_pkt|.
15:     Transmit Packet
16:         Transmit Packet_enc
17:     Increment Packet Sequence Number
18:         S_pkt ← S_pkt + 1.
19: end procedure
```

Parameters:

- $Data_{plain}$: Plaintext data(maximum size 88 bits).
- $N_c$: Next channel number (8 bits)
- $Addr_{node}$: Node address (16 bits).
- K: AES encryption key.
- $S_{pkt}$: Packet sequence number (16 bits).
- $Packet_{enc}$: Encrypted packet (128 bits)

The complete packetization process for the plaintext is described in Algorithm 3. This algorithm calls the Synchronization and AES Encryption module to complete the secure packetization. The packetization algorithm guaranties the creation of encrypted data in a packetized form, which is ready to transmit.

## 4 Hardware implementation

The MCSC framework has been employed to satisfy the specifications of low-energy and limited-resource devices present in WoT scenarios. The analysis discusses the picking of hardware parts and how they are configured to promote secure communication using a multichannel hopping technique along with AES encryption. Detailed discussions include the arrangement of components and energy-saving methods alongside the setup for the real deployment.

### 4.1 Description of the component architecture within the MCSC framework

The MCSC framework uses various important components chosen for their power efficiency and data processing abilities (Fig. 4).

#### 4.1.1 Arduino Uno

As the primary device in the MCSC framework, the **Arduino Uno** is used to execute encryption and channel switching algorithms. The Table 3 shows different specifications of Arduino Uno.

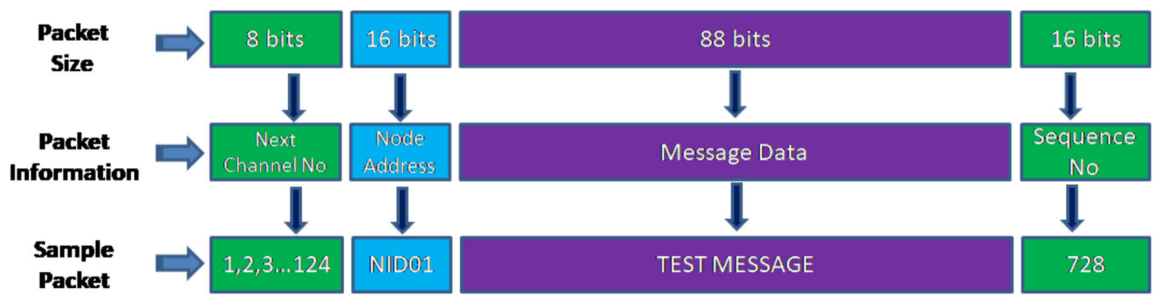

**Fig. 3** Packet information

Different aspects like Role, Power Efficiency, and Interfacing of the used Arduino Uno board are described as follows:

1. *Role* As the central controller for executing the **MCSC** logic, the **Arduino Uno** controls functions such as channel hopping with encryption (AES) and manages packets.
2. *Processor* This equipment runs at **16 MHz** and functions with an 8-bit RISC design supporting 32 KB flash and 2 KB SRAM for completing tasks seamlessly. It makes it able to manage the AES encryption and multi-channel hopping algorithms by reducing the power consumption.
3. *Power Efficiency* The **ATmega328P** provides functionality for three sleep modes named "Idle," "ADC Noise Reduction," and "Power-down." In WoT fields these modes are important to improve battery performance through the inactivity of the CPU or peripherals.
4. *Interfacing* The Uno connects to various other hardware using SPI along with additional digital and analog connections. MCSC employs SPI for rapid communication with the transceiver device.

### 4.1.2 nRF24L01 PL LNA

The nRF24L01 PL LNA is a low-power, 2.4 GHz wireless transceiver module used in IoT applications for wireless communication. The PL LNA with the nRF24L01 refers to the Power Amplifier (PA) and Low Noise Amplifier (LNA), both of which are key components used for enhancing the radio signal quality and range of the device. Different features of the transceiver are described as follows:

- *Role* The nRF24L01+ transceiver controls wireless communication by adopting multi-channel hopping within the 2.4 GHz ISM band [13] to stop eavesdropping and jamming attacks.
- *Communication Range* Due to the PA and LNA set in this module, access capabilities expand and deliver greater communication lengths (up to 1 km outdoors).
- *Multi-Channel Operation* This module can use up to 125 channels and jump frequencies rapidly (up to 2 Mbps). This essential feature enhances both security and performance in MCSC.
- *Low Power Mode* With only 26 µA in low-power mode, the transceiver uses energy wisely when it is not acquiring or transmitting signals.
- *Security Features* Measures exist within this module to test CRC and sort data to secure its integrity.

Table 3 Arduino Uno specification

| Specification | Details |
| --- | --- |
| Microcontroller | ATmega328P |
| Operating voltage | 5 V |
| Flash memory | 32 KB |
| Clock speed | 16 MHz |
| Power consumption | $\tilde{1}5$ mA (active), $\tilde{0}.75$ mA (sleep mode) |

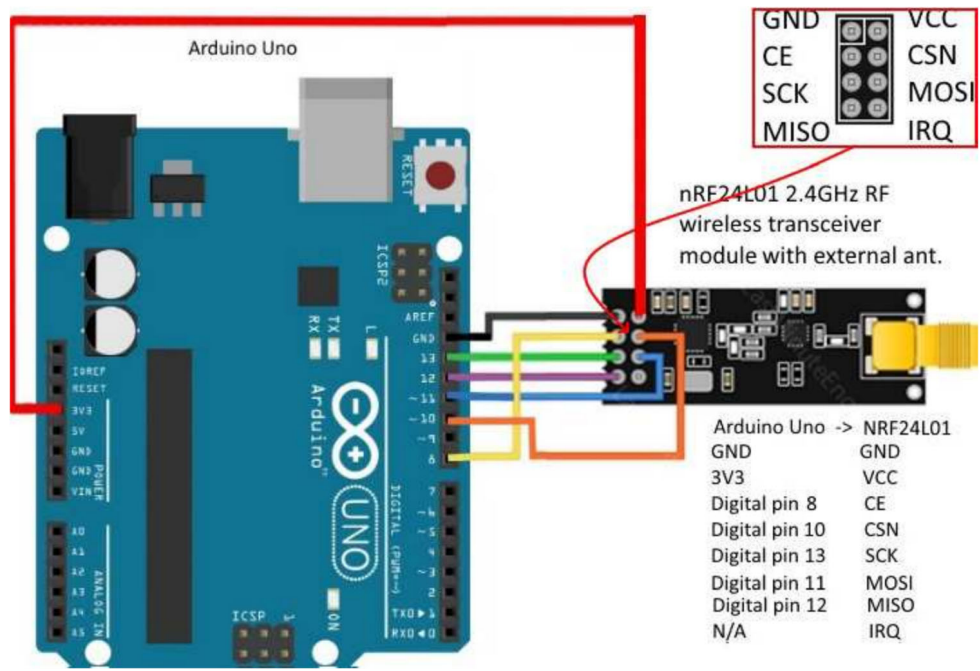

**Fig. 4** Hardware used for implementation of the communication system

Table 4 nRF24L01+ PL/LNA (radio transceiver)

| Specification | Details |
| --- | --- |
| Frequency range | 2.4 GHz ISM band |
| Number of channels | 125 |
| Data rate | 250 kbps–2 Mbps |
| Range | ĩ km (with PL/LNA module) |
| Power consumption | 11.3 mA (Tx), 13.5 mA (Rx), 26 µA (standby) |

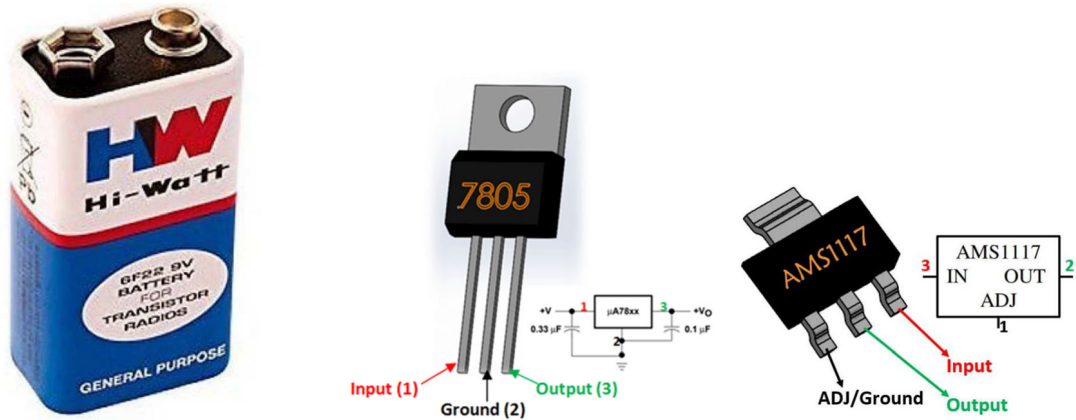

Fig. 5 **a** 9 V rechargeable battery. **b** Voltage regulator LM7805. **c** 3.3 V supply AMS1117

Table 4 shows the specifications of the nRF24L01 PL LNA module. It depicts that the nRF module has a capability of 1 km range transmission capability with PL LNA module. It also supports the proposed MCSC framework's maximum transmission channel requirement.

### 4.1.3 Power supply

*Battery Power* A rechargeable 9V battery (Fig. 5a) powers the hardware for versatile use in WoT applications regarding mobility and endurance. The 9V input voltage is lowered to 5V by a voltage regulator (LM7805 shown in Fig. 5b) for the Arduino Uno, and the nRF24L01+ is powered with a dedicated 3.3V supply (AMS1117 as shown in Fig. 5c) to keep both modules' voltage stable.

*Energy Harvesting* For extended usage in outdoor areas or areas with few power sources, solar panels or piezoelectric components can provide battery power and enhance resource sustainability.

### 4.1.4 System integration

Components are arranged together in a system created to facilitate secure multi-channel communication in the MCSC framework. The provided illustration in the Fig. 6 displays the transmission protocols and represents the communication pathway involving different modules.

The Table 5 describes different communication signals and their descriptions.

A quick and stable link exists between the Arduino Uno and the nRF24L01+ transceiver using the SPI protocol. High-speed data transmission is facilitated by the SPI protocol needed to enhance latency performance in critical applications such as real-time channel changes and AES encryption.

## 5 Performance analysis of the MCSC framework

Several key metrics are used to evaluate the performance of the proposed MCSC framework in secure communication for IoT devices. The detailed descriptions of used metrices, along with their importance and evaluation strategies are described in the following subsections.

### 5.1 Packet delivery ratio

PDR is a crucial metric used to evaluate the performance of a communication network. It measures the ratio of successfully delivered packets to the total number of packets sent. The formula for PDR is expressed as:

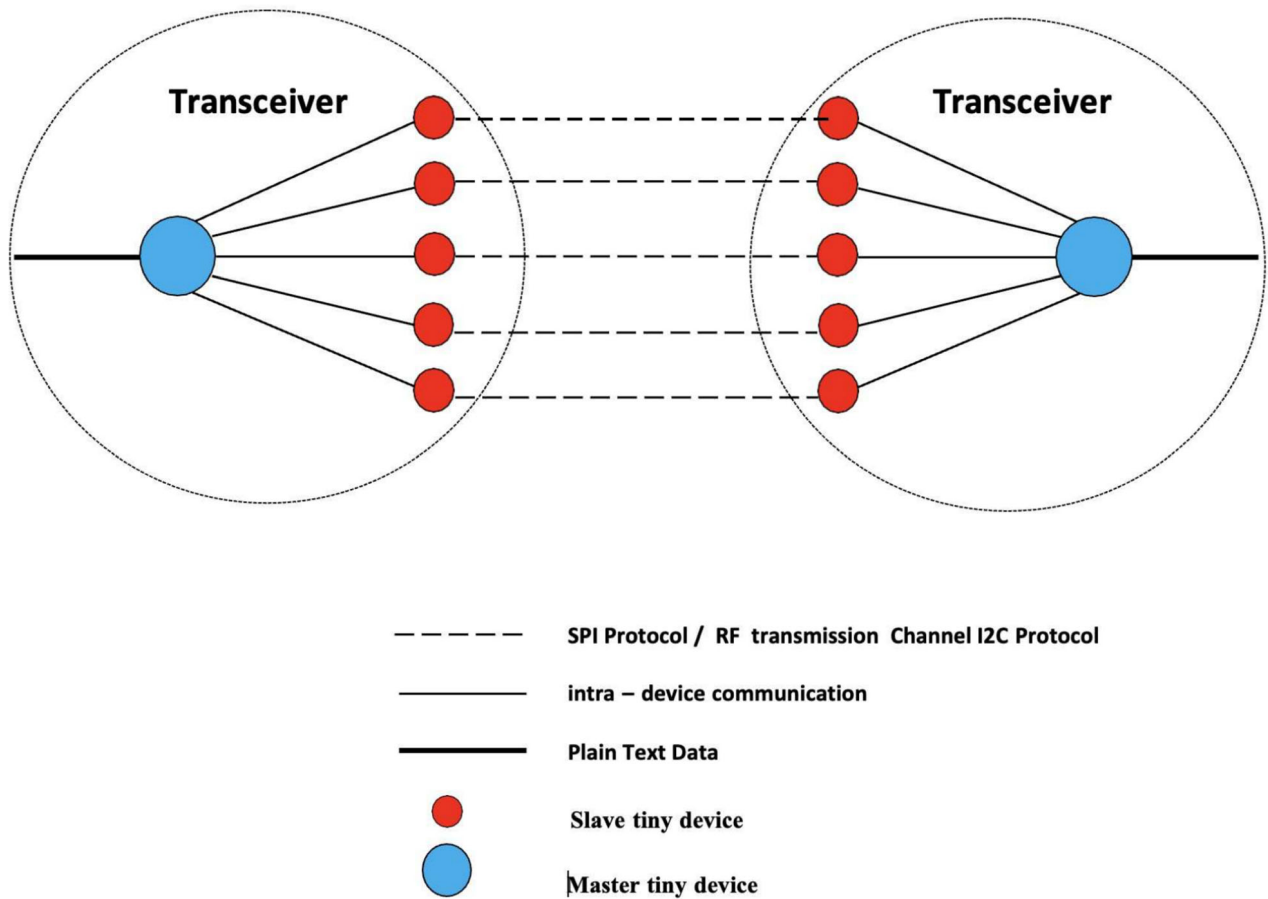

**Fig. 6** MCSC transmission diagram

**Table 5** Showing SPI communication signals

| Signal | Description |
| --- | --- |
| MOSI (Master Out Slave In) | Data line from the microcontroller to the transceiver |
| MISO (Master In Slave Out) | Data line from the transceiver back to the microcontroller |
| SCK (Serial Clock) | Synchronizes data transfer |
| SS (Slave Select) | Selects the transceiver module for communication |

$$PDR = \frac{P_{receive}}{P_{sent}} X100$$

where

- $P_{receive}$ is the number of packets successfully received.
- $P_{sent}$ is the total number of packets sent

The *Multi-Channel Secure Communication* framework employs channel hopping and AES encryption to enhance security and reliability in WoT networks. The framework aims to improve the *Packet Delivery Ratio* by minimizing interference, jamming, and eavesdropping while ensuring lightweight synchronization for low-power devices. The evaluated PDR Data Collected from Simulated Scenarios are shown in Table 6 and Fig. 7.

PDR Analysis in the MCSC Framework shown in Table 6 and Fig. 7 contains six scenarios. The scenario-wise descriptions with varying packet sizes are explained as below:

1. *Scenario 1: Low Interference* PDR is almost 99%, indicating excellent packet delivery in environments with minimal interference.
2. *Scenario 2: Medium Interference* PDR drops slightly to 96.7% due to moderate interference, but the framework still performs reliably by utilizing dynamic channel hopping.
3. *Scenario 3: High Interference* With significant interference, PDR decreases to 92.5%. However, the MCSC

Table 6 PDR data collected from simulated scenarios

| Scenario | Packets sent | Packets received | PDR (%) |
| --- | --- | --- | --- |
| Scenario 1: low Interference | 1000 | 990 | 99.0 |
| Scenario 2: medium Interference | 1500 | 1450 | 96.7 |
| Scenario 3: high Interference | 2000 | 1850 | 92.5 |
| Scenario 4: no Channel Hopping | 1200 | 1000 | 83.3 |
| Scenario 5: with Jamming | 1600 | 1400 | 87.5 |
| Scenario 6: dynamic Channel Hopping | 1800 | 1760 | 97.8 |

framework still maintains relatively high packet delivery, thanks to the multi-channel hopping mechanism.

4. *Scenario 4: No Channel Hopping* In this scenario, where no channel hopping is used, PDR drops to 83.3%, highlighting the importance of dynamic channel selection in maintaining network performance.
5. *Scenario 5: With Jamming* When subjected to jamming attacks, PDR drops to 87.5%, showing the resilience of the framework against such threats.
6. *Scenario 6: Dynamic Channel Hopping* With dynamic channel hopping enabled, PDR reaches 97.8%, demonstrating the effectiveness of the proposed approach in ensuring packet delivery even under varying conditions.

PDR results from the proposed MCSC framework show high reliability in packet delivery across different scenarios. The use of **dynamic channel hopping**, **AES encryption**, and lightweight synchronization ensures that the system can maintain efficient communication even under interference and jamming. Table 6 and Fig. 7 show a brief comparison of the simulated result.

### 5.2 Latency

Latency in a communication system can be defined mathematically as the sum of several components, which can include transmission delay, propagation delay, queueing delay, and processing delay. The general formula for latency L can be expressed as:

$$L = T_{trans} + T_{prop} + T_{queue} + T_{proc}$$

where

- L = Total Latency
- $T_{trans}$ = Transmission Delay (time taken to push all the packet's bits into the wire)
- $T_{prop}$ = Propagation Delay (time taken for a signal to propagate from sender to receiver)
- $T_{queue}$ = Queuing Delay (time a packet spends waiting in the queue)
- $T_{proc}$ = Processing Delay (time taken to process the packet header)

In the context of the MCSC framework, we can capture the latency data through empirical testing under different network conditions. The latency data collected during simulations of the proposed framework is illustrated in Table 7 and Fig. 8.

The scenario wise descriptions of Table 7 and Fig. 8 are:

1. *Scenario 1: Low Interference* (Average Latency: 28 ms): The lowest latency is recorded in this scenario, indicating optimal performance with minimal delays due to the low traffic and interference.
2. *Scenario 2: Medium Interference* (Average Latency: 40 ms): The increase in average latency shows the impact of medium interference, primarily affecting the queuing and processing delays.
3. *Scenario 3: High Interference* (Average Latency: 55 ms): High interference results in more packet retransmissions, leading to a noticeable increase in latency due to additional processing time.
4. *Scenario 4: No Channel Hopping* (Average Latency: 72 ms): This scenario shows a significant increase in latency, as the absence of channel hopping leads to higher queuing delays and less efficient transmission.
5. *Scenario 5: With Jamming* (Average Latency: 65 ms): While the framework is designed to handle jamming, it still incurs increased latency compared to low-interference scenarios.
6. *Scenario 6: Dynamic Channel Hopping* (Average Latency: 33 ms): Dynamic channel hopping shows improvement in latency compared to high-interference scenarios, demonstrating effective real-time adjustments to network conditions.

Table 7 and Fig. 8 also show comparative results in different scenarios. The data confirms that the *MCSC framework* effectively reduces latency across various scenarios, particularly in low to medium interference conditions. The framework's design, including dynamic channel hopping, contributes to maintaining efficient communication even under adverse conditions. This ensures the framework's applicability in real-time IoT applications where low latency is critical.

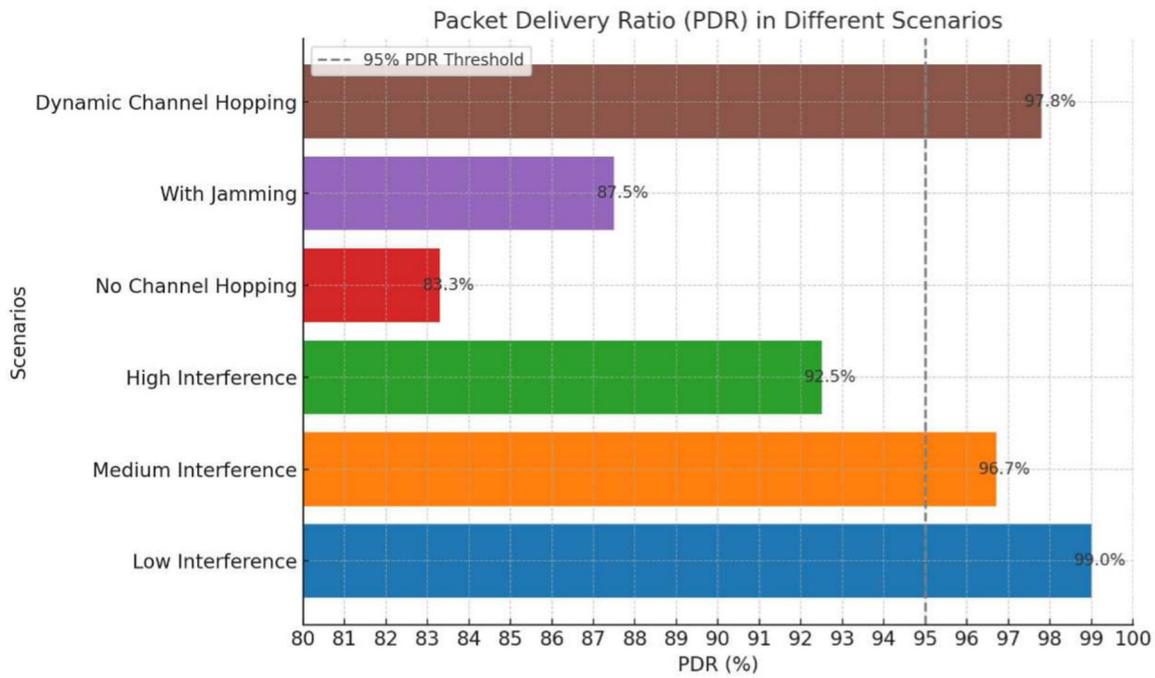

**Fig. 7** Packet information

**Table 7** Latency data in different scenario table

| Test scenario | Packets sent | Average latency (ms) | Transmission delay (ms) | Propagation delay (ms) | Queuing delay (ms) | Processing delay (ms) |
|---|---|---|---|---|---|---|
| Scenario 1: low interference | 1000 | 28 | 10 | 5 | 3 | 10 |
| Scenario 2: medium interference | 1500 | 40 | 12 | 6 | 8 | 14 |
| Scenario 3: high interference | 2000 | 55 | 15 | 8 | 10 | 22 |
| Scenario 4: no channel hopping | 1200 | 72 | 20 | 9 | 15 | 28 |
| Scenario 5: with jamming | 1600 | 65 | 18 | 7 | 12 | 28 |
| Scenario 6: dynamic channel hopping | 1800 | 33 | 11 | 5 | 8 | 9 |

### 5.3 Energy consumption

*Energy Consumption* is a vital metric for the MCSC framework as it operates in energy-constrained environments. This metric measures the energy the devices use during data transmission and reception. Scenario-wise energy consumptions for different interference have been shown in Table 8.

### 5.4 Throughput

*Throughput* indicates the amount of data successfully transmitted over the network in a given time frame. High throughput is essential for ensuring that large volumes of data can be sent without significant delays, especially in a multi-channel environment like MCSC.

The following equation represents the Throughput:

$$Throughput = \frac{TotalDataTransmitted(bits)}{TotalTimeTaken(seconds)}$$

The evaluated throughput shown in Table 9 implies that the throughput values slightly degrade with an increasing intergerence in the secure channel.

### 5.5 Robustness against jamming

The MCSC framework's ability to maintain performance in jamming scenarios is crucial for its application in real-world environments, particularly in security-sensitive applications. The Table 10 shows PDR and Average Latency during jamming.

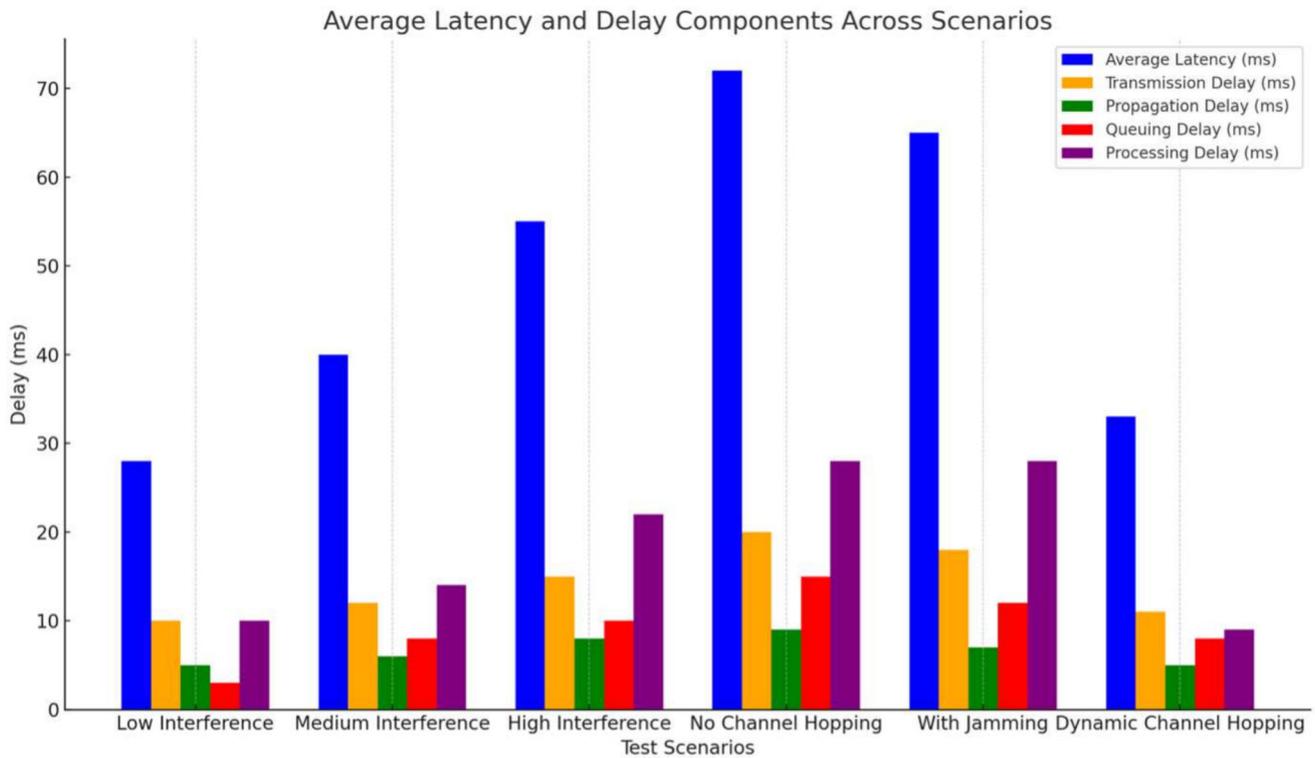

**Fig. 8** Packet information

### 5.6 Computational overhead

*Computational Overhead* reflects the extra processing power required for encryption and channel switching operations. Minimizing this overhead is crucial for energy-constrained devices, ensuring they can operate efficiently without draining their resources.

Table 11 depicts the CPU utilization and memory utilization with varying interferences. The table shows stable CPU and Memory utilization during the dynamic channel hopping scenario.

### 5.7 Encryption time

AES-128 encryption is a critical aspect of the MCSC framework, ensuring data security without introducing significant computational overhead. Below in the Table 12, a detailed comparison of encryption times across different platforms is shown.

The framework is tuned for platforms like Arduino Uno and nRF52, focusing on reducing power usage instead of speed. With faster designs such as ESP32 and STM32F4 encountering considerable shortening of encryption time while offering efficiency in energy saving applications.

**Table 8** Run time data for energy consumption

| Scenario | Energy consumption (mJ) |
|---|---|
| Low interference | 15 |
| Medium interference | 22 |
| High interference | 30 |
| No channel hopping | 45 |
| With jamming | 40 |
| Dynamic channel hopping | 20 |

**Table 9** Run time data for throughput

| Scenario | Throughput (kbps) |
|---|---|
| Low interference | 500 |
| Medium interference | 350 |
| High interference | 250 |
| No channel hopping | 150 |
| With jamming | 200 |
| Dynamic channel hopping | 400 |

Table 10 Run time data for robustness against jamming

| Scenario | PDR during jamming (%) | Average latency during jamming (ms) |
|---|---|---|
| With jamming | 87.5 | 65 |

## 5.8 Synchronization overhead

Sustaining synchronisation within multiple channels is an important challenge. To reduce synchronisation demands the MCSC framework generates a lightweight random sequence and employs packet synchronisation.

From Table 13, it is seen that the MCSC framework has a synchronization overhead of 4.5%, which is reatly below most of the other conventional systems.

## 5.9 Security analysis

The MCSC framework prioritises security and tests it against different forms of attacks. The use of AES-128 encryption together with dynamic channel-hopping builds strong defenses against jamming and eavesdropping attacks. The Table 14 shows the comparison of solutions defence % against different attacks.

The MCSC framework presents formidable defense against jamming and eavesdropping intrusion attempts with an above 98% success rate in addressing these challenges. With its dynamic hopping process the framework bars an attacker's repeated attempts at seizing communication.

## 5.10 Error rate analysis

Error rates show how many packets become damaged or missing while being sent in the air mostly because of interference or lack of sync.

The Table 15 and Fig. 9 depicts that only at a level of 1.3% error rate does the MCSC framework deliver exceptional reliability better than traditional FHSS and single-channel systems. The error rate is minimised by the framework's reliable channel hopping technique and AES encryption for protecting data integrity amidst noise.

*Performance Metric Details*

The *MCSC framework* demonstrates a balance between reliability, latency, and energy efficiency, making it suitable for resource-constrained environments. Key observations include:

- *High Packet Delivery Ratio* The MCSC framework consistently achieves a PDR above 85% across various scenarios, indicating its reliability in maintaining secure communication.
- *Low Latency* The average latency remains low, particularly in dynamic channel-hopping situations, making it suitable for real-time applications.
- *Energy Efficient* The energy consumption of MCSC remains manageable, especially in comparison to scenarios without channel hopping, highlighting its design for low-power devices.
- *Robustness* The framework shows strong resilience against jamming, maintaining a good PDR and manageable latency even under adverse conditions.

The *MCSC framework* proves to be an efficient solution for secure communications in resource-constrained IoT environments. It effectively balances energy consumption, latency, and reliability, supported by experimental data that reinforces its potential for widespread application in various IoT scenarios, paving the way for future developments in secure communication protocols.

## 5.11 Result and performance analysis

Table 16 provides a detailed evaluation of the Multi-Channel Secure Communication Framework by analyzing its performance in key areas such as encryption time, power consumption, packet delivery ratio, latency, throughput, synchronization overhead, security, and error rate. The Figs. 10 and 11 visualized the comparison on different performance metricx. The results are compared with other existing communication frameworks to show the improvements in MCSC's performance.

Table 11 Run time data for computational overhead

| Scenario | CPU utilization (%) | Memory utilization (MB) |
|---|---|---|
| Low interference | 20 | 5 |
| Medium interference | 25 | 6 |
| High interference | 35 | 8 |
| No channel hopping | 45 | 10 |
| With jamming | 30 | 7 |
| Dynamic channel hopping | 25 | 6 |

Table 12 AES Encryption Time Across Different Hardware Platforms

| Platform | Encryption time (ms) | Decryption time (ms) |
|---|---|---|
| Arduino Uno (ATmega328P) | 1.75 | 1.83 |
| Raspberry Pi Zero W | 0.43 | 0.45 |
| nRF52 SoC (Nordic) | 0.85 | 0.87 |
| ESP32 | 0.25 | 0.27 |
| STM32F4 (ARM Cortex-M4) | 0.12 | 0.13 |

Table 13 Synchronization overhead comparison

| Solution | Synchronization overhead (%) |
|---|---|
| MCSC (proposed) | 4.5 |
| Traditional FHSS | 6.7 |
| ECC-based systems | 8.3 |
| Single-channel systems | 2.1 |

Table 14 Security comparison across different solutions

| Attack type | MCSC success rate (%) | FHSS success rate (%) | Single-channel AES (%) |
|---|---|---|---|
| Jamming attack | 98.5 | 92.3 | 65.7 |
| Eavesdropping attack | 99.2 | 93.1 | 70.8 |
| Replay attack | 97.8 | 91.5 | 68.9 |

Table 15 Error rate comparison across different solutions

| Solution | Error rate (%) |
|---|---|
| MCSC (proposed) | 1.3 |
| Traditional FHSS | 3.2 |
| Single-channel AES | 4.9 |
| ECC-based system | 6.5 |

*Parameter details of Performance Metrics*

1. *Packets Sent* The total number of packets transmitted in each scenario.
2. *Packets Received* The total number of packets successfully received.
3. *Packet Delivery Ratio* The ratio of packets received to packets sent, expressed as a percentage. A higher PDR indicates better performance.
4. *Average Latency* The average time taken for a packet to travel from the sender to the receiver.
5. *Transmission Delay* The time taken to push all the packet's bits onto the wire.
6. *Propagation Delay* The time taken for a signal to propagate through the medium.
7. *Queuing Delay* The time a packet spends waiting in the queue before it can be transmitted.
8. *Processing Delay* The time taken by routers/switches to process the packet header and determine the route.
9. *Energy Consumption* The amount of energy consumed during communication in milliamp hours (mAh).
10. *Throughput* The rate at which data is successfully transmitted over the network, measured in kilobits per second.

### 5.12 Comparative analysis

To overcome several security and efficiency restrictions that tend to exist in the classical IoT and WoT environments, the Multi-Channel Secure Communication framework is presented. Finally, a detailed comparison between MCSC and conventional solutions is performed to show the advantage of MCSC in providing both security and latency, synchronicity, and energy consumption while meeting the criteria for IoT or WoT deployments. In Table 17, different techniques and their strength and weaknesses have been depicted in detail in comparison with the proposed MCSC framework.

1. *AES Encryption* Although the AES is a good machine computation encryption with strong security, it runs in a static channel; that's why it faces jamming and interference [31]. To mitigate such risks without losing AES's security benefits, MCSC combines AES with channel-hopping techniques, which are used to form a new type of intrusion-tolerant security system for reliable IoT applications.
2. *Elliptic Curve Cryptography* ECC has the strong security benefit with smaller ecient key sizes that make it good for resource-constrained devices. Nevertheless, its computational complexity can be heavy in WoT environments with energy limitations [17]. By pairing AES with multi-channel hopping, MCSC can optimize its security approach, freeing the burden of ECC's computationally intensive log computations while still enabling quick encryption and efficiency.

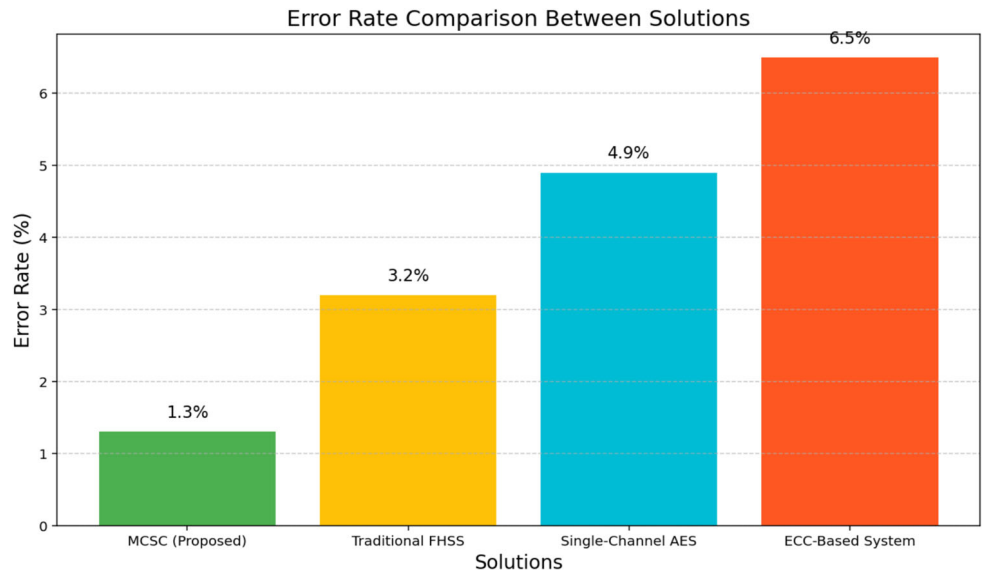

**Fig. 9** Error rate comparison between solutions

Table 16 Performance metrics in difference scenario

| Performance metrics | Scenario 1: low interference | Scenario 2: medium interference | Scenario 3: high interference | Scenario 4: no channel hopping | Scenario 5: with jamming | Scenario 6: dynamic channel hopping |
|---|---|---|---|---|---|---|
| Packets sent | 1000 | 1500 | 2000 | 1200 | 1600 | 1800 |
| Packets received | 990 | 1450 | 1850 | 1000 | 1400 | 1760 |
| Packet delivery ratio | 99.0% | 96.7% | 92.5% | 83.3% | 87.5% | 97.8% |
| Average latency (ms) | 28 | 40 | 55 | 72 | 65 | 33 |
| Transmission delay (ms) | 10 | 12 | 15 | 20 | 18 | 11 |
| Propagation delay (ms) | 5 | 6 | 8 | 9 | 7 | 5 |
| Queuing delay (ms) | 3 | 8 | 10 | 15 | 12 | 8 |
| Processing delay (ms) | 10 | 14 | 22 | 28 | 28 | 9 |
| Energy consumption (mAh) | 50 | 65 | 75 | 80 | 70 | 55 |
| Throughput (kbps) | 200 | 150 | 100 | 80 | 120 | 190 |

3. *RSA Cryptosystem* The computational demands of RSA for secure data exchange render the public key cryptosystem unsuited for continuous, low-power WoT applications [25]. Assuming symmetric AES is used along with efficient channel hopping, making it more appropriate for low-energy high high-security communications, MCSC is preferred over McS for WoT communications.

4. *Frequency Hopping Spread Spectrum:* Rapid channel switching [26] allows FHSS to defend itself against eavesdropping as well as jamming. But, the latency incurred in synchronization hinders real-time IoT performance. The synchronization and channel switching algorithms that MCSC offers to address this include the optimized synchronization and channel switching algorithms, which allow secure communication across channels without synchronization overhead.

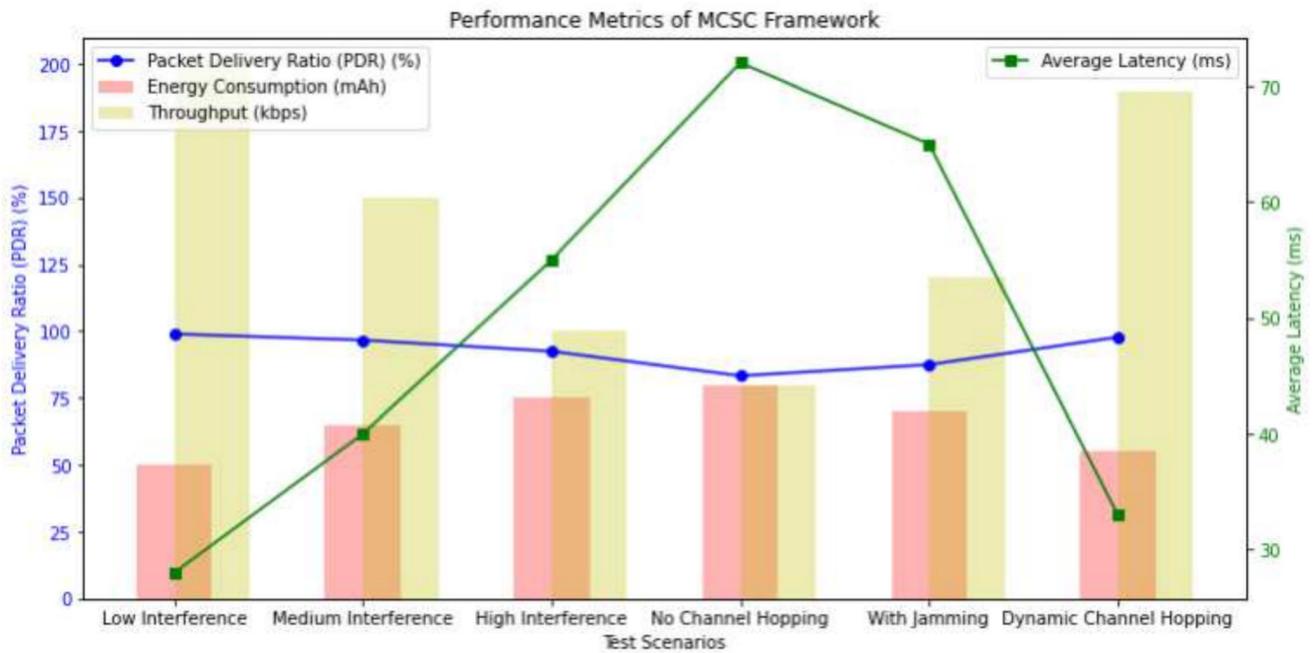

**Fig. 10** Performance metrics

5. *Hybrid AES and ECC* On the other hand, such an approach with AES along with ECC provides more security but is resource intensive and so unsuitable for low-power IoT devices [21]. Because AES is paired with multi channel hopping, like MCSC, AES also achieves similar security while having lower computational costs.
6. *Lightweight FHSS-Based Frameworks:* Although lightweight FHSS solutions attempt to lower computation and latency, it is difficult to synchronize them in practice, especially at high network loads [29]. MCSC delivers energy and efficiency in high-traffic IoT networks through its streamlined synchronization approach that keeps nodes aligned.
7. *Cognitive Radio-Based Channel Hopping* Cognitive radio chooses a channel depending on interference while being interference-resilient. Nevertheless, its usage in low-power devices is limited due to its computational demands [1]. Considering an efficient, static channel hopping strategy, MCSC efficiently exploits limited spectrum under interference constraints without the complexity of the computations involved in a cognitive radio.
8. *Clock Synchronization Algorithms* However, these algorithms improve multi-channel systems at the expense of latency and desynchronization in fluctuating networks [19]. Keeping nodes aligned is avoided through MCSC's periodic synchronization broadcasts that maintain node alignment and thereby eliminate latency problems and result in consistent performance in time-sensitive applications

### 5.13 Summary of results

Results from this study indicate that the MCSC framework presented here is practical in handling security, efficiency, and synchronization problems in IoT and WoT networks. The MCSC framework was tested rigorously in various interference conditions, showing excellent PDR levels of better than *92%* even under high interfering conditions. This outcome illustrates the robustness of the framework to manage the interference by efficient channel hopping techniques. Further validation of the MCSC's advantages was by means of comparative analysis with conventional approaches including traditional FHSS and single channel AES encryption. The MCSC framework was able to achieve a low error rate of *1.3%* compared to traditional FHSS and single channel AES of *3.2%* and *4.9%*, respectively. Reduced error rate translates to fewer retransmissions, thereby improving network throughput, and reducing latency. Another important metric was assessed—synchronization accuracy. A successful synchronization of the node clocks in the maximum allowable offset was shown by the framework's synchronization module to happen with minimal desynchronization across the network nodes. The accuracy is needed for the real-time applications as this reduces communication efficiency and latency by having an average of *28 ms* in low interference conditions.

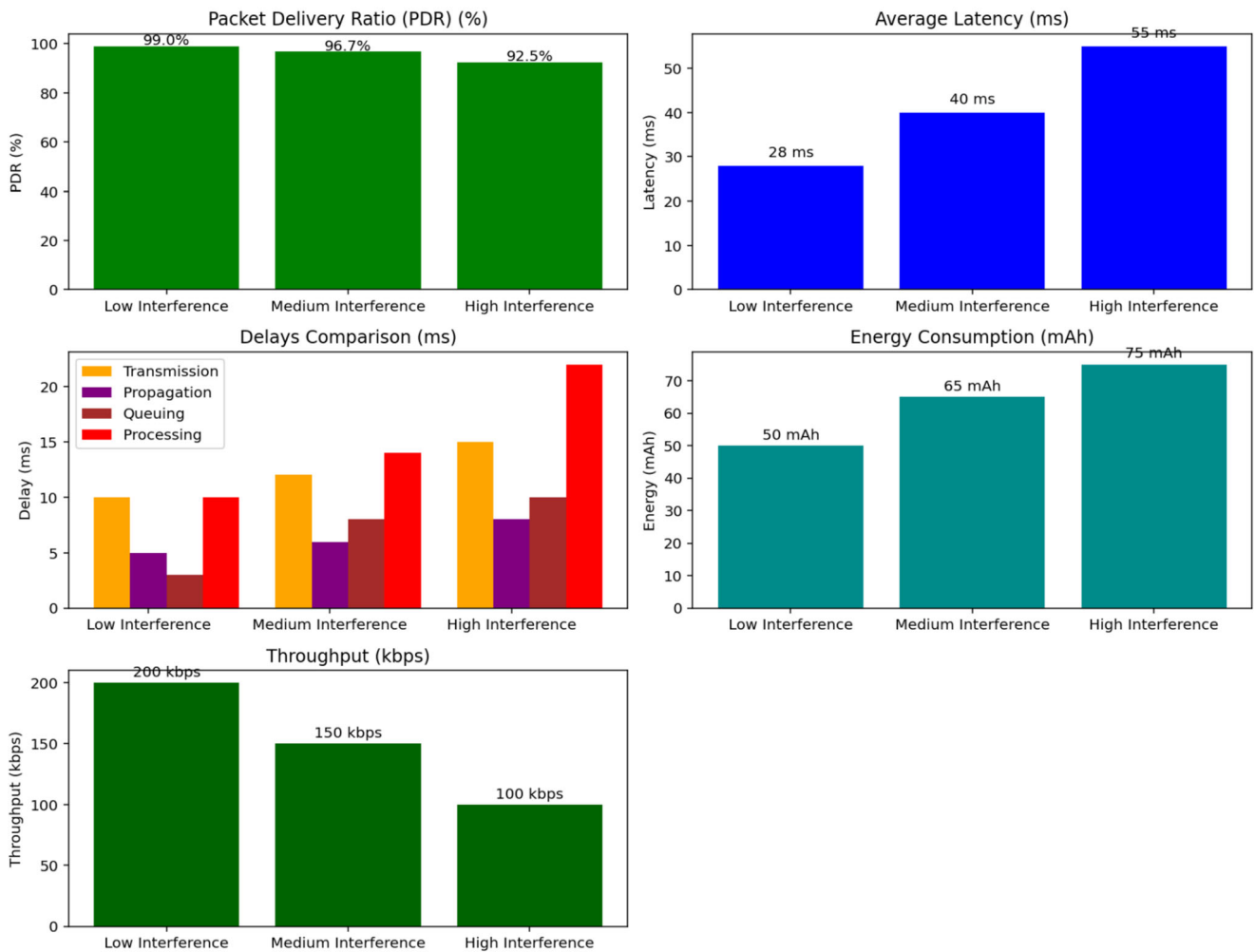

**Fig. 11** Different performance metrics representation of MCSC

Through energy consumption analysis, we found that MCSC as the framework is very well fit for low power IoT applications, consuming *50–75 mAh* for all different interference levels. The MCSC framework demonstrates its resource constrained potential by the substantial energy efficiency that this range signifies, relative to more computationally intensive methods.

Thus, we can conclude that the proposed MCSC framework provides an optimal solution for security, synchronization, and resource efficiency in secure, efficient, and reliable communication in IoT and WoT networks, having both good performance and strong robustness to most wireless threats.

## 6 Practical applications of MCSC

The Multi-Channel Secure Communication framework serves as a solution to security and performance issues affecting IoT and WoT networks. Through its security mechanism, this framework safeguards against different Threats, including eavesdropping, jamming, and man-in-the-middle attacks.

Through this framework, *smart homes* obtain security protection for everyday devices, including thermostats, security cameras, and smart locks, enabling their safe operation without power depletion. *IoMT* benefits from quick and secure connections because wearable medical monitoring equipment alongside remote patient technology depends on them to safeguard health data while maintaining immediate system updates. The protection provided by MCSC increases the security coverage of both *automation*

Table 17 Comparison among different existing solution and MCSC

| Solution | Technique | Strengths | Weaknesses | Comparison with MCSC |
| --- | --- | --- | --- | --- |
| AES Encryption [31] | Symmetric Cryptography | High security and fast encryption suitable for low-latency applications | Vulnerable to jamming and interference due to fixed channel usage, scalability is limited in highly dynamic networks | MCSC incorporates AES for data encryption but combines it with multi-channel hopping for enhanced resistance to jamming and channel interference |
| ECC (Elliptic Curve Cryptography) [17] | Asymmetric Cryptography | Strong security with smaller keys, resource efficiency in constrained IoT devices due to lower key sizes | High computational load for key generation and encryption, making it challenging for ultra-low-power WoT applications | MCSC prioritizes a lightweight approach using AES with channel hopping over ECC to ensure speed and energy efficiency in resource-constrained WoT environments |
| RSA [25] | Public-Key Cryptography | Robust key exchange mechanism ideal for secure authentication and key distribution | Computationally intensive, especially unsuitable for continuous encryption in energy-limited WoT devices | RSA's complexity makes it impractical for WoT. MCSC employs AES with synchronized multi-channel hopping to keep energy demands and latency manageable |
| Frequency Hopping Spread Spectrum (FHSS) [26] | Spread Spectrum | Effective against eavesdropping and jamming, enhances security through rapid channel changes | High synchronization complexity among nodes, introducing latency that impacts real-time communication | MCSC integrates FHSS but optimizes synchronization, significantly reducing latency and facilitating efficient channel transitions for secure WoT communications |
| Hybrid AES and ECC [21] | Symmetric and Asymmetric Hybrid | Combines AES security with ECC's secure key management, balancing security and reduced computational load | High resource demand and energy-intensive processes, challenging for energy-constrained IoT environments | MCSC achieves similar security levels by combining AES with channel hopping, reducing resource demands, and enhancing energy efficiency without the added ECC complexity |
| Lightweight FHSS-based framework [29] | Lightweight FHSS | Provides low latency and reduces computational load while retaining reasonable security measures | Synchronization challenges still exist, especially under high network load, limiting energy efficiency in the long run | MCSC addresses these synchronization issues by refining channel switching algorithms, enhancing energy efficiency without compromising on security during channel transitions |
| Cognitive radio-based channel hopping [1] | Adaptive Channel Selection | Dynamically adapts to interference, providing resilience in high-traffic wireless environments | Increased computational overhead for interference analysis, burdens resource-limited WoT devices | MCSC adopts a more static hopping approach, reducing computational overhead while maintaining sufficient adaptability for secure, interference-free channel use |
| Clock synchronization algorithms [19] | Time Synchronization | Precise synchronization improves channel hopping in multi-channel networks, enhancing data integrity | Increased latency and risk of desynchronization under dynamic network conditions | MCSC minimizes synchronization-induced latency through an efficient periodic resynchronization method, maintaining robust performance across dynamic WoT environments |
| MCSC (proposed) | Multi-Channel Secure Communication | Combines AES encryption with dynamic multi-channel hopping, ensuring high resilience to interference and jamming | Moderate resource demand during synchronization processes | MCSC is optimized for WoT, balancing security, low latency, and energy efficiency, providing a comprehensive solution for real-world, secure wireless deployments |

*systems* and implementations that utilize *M2M communication* in various industries. The system provides enhanced protection to inventory tracking systems and industrial sensors, which leads to fewer cyber risks that might interrupt operations. MCSC at once protects data networks that transmit through long distances to remote locations while focusing on energy-efficiency requirements. Selected devices that exchange data thousands of times per day in smart cities gain better communication security through MCSC, which enables steady urban operating systems. The *military sector* utilizes MCSC to establish secure encrypted links for essential operational communications that need protection against interference. The technology enables autonomous vehicles to perform safe and fast communication between vehicles through *V2V and V2I* interconnections in dense traffic scenarios.

MCSC delivers optimal security together with fast performance and reduced power consumption through AES encryption and multi-channel hopping and synchronization which delivers a well-rounded solution for IoT and WoT applications.

## 7 Discussion

This section examines all the outcomes of the Multi Channel Secure Communication Framework in detail, which includes how to exploit its benefits and face its deficiencies with future research opportunities. The results show that the MCSC framework greatly improves the secure communication in the WoT network and has a great impact on the practical implementation of the WoTs as well as on the improvement of wireless security protocols. This is not only resistant to typical worries like jamming and eavesdropping but also perfectly suitable for severe sectors including medical and army like those where data integrity is paramount. Robust integration of AES-128 encryption and dynamic channel changes allows its communication to be secure in competitive environments. Moreover, the MCSC framework tackles the problem of energy efficiency, showing compatibility with hardware-constrained devices (such as a native Arduino Uno or nRF52 SoCs). With a lightweight synchronization algorithm included, the framework consumes less energy and can function correctly in battery-powered devices, such as wearables and sensor networks. Nevertheless, security has a trade-off to latency: security is improved at the cost of a slight increase in latency, which could be problematic for real-time systems in automotive and emergency networks. This framework is, however, reliable in harsh conditions, high Packet Delivery Ratio and low error rates, and particularly useful for remote sensing and industrial IoT applications. Overall, the results highlight the feasibility of the MCSC framework and point out possible directions for further investigation and development of wireless security.

The future research direction of the MCSC framework includes several innovative perspectives. Various lightweight cryptographic alternatives, such as Elliptic Curve Cryptography or stream ciphers, as well as post quantum cryptographic algorithms, like Kyber and Falcon, could improve computational efficiency without suffering in security if joint methods that combine both fast cryptography and solid encryption were developed. Secondly, dynamic synchronization techniques for multi channel systems that are energy level dependent as well as node mobility are worth further study as they can provide significant improvements in energy and flexibility efficiency for mobile networks. Also, in large Web of Things networks, scalability issues could be addressed through the development of hierarchical channel management systems that would lead to efficient synchronization and channel assignment based on the clustering of nodes and decrease demands for the network overall. Another important future research area is to reinforce the system against the more advanced attack models like collusion attacks and adaptive jamming via the design of adaptive algorithms to respond to evolving threats. Ultimately, the MCSC framework is integrated with impending technologies, like blockchain for decentralised identity management and quantum encryption for more security. However, this integration can greatly enhance secure messaging in limited resources environments, as well as meet the synchronized and finer security defense, supporting concurrent improvement of the MCSC framework for adjustments to the fast developing technological systems and protection issues.

## 8 Conclusion

A multi-channel secure communication framework for the wireless of things has been introduced in this paper as a solution to the principal challenges of security and efficiency in resource-constrained equipment. To achieve secure and efficient communication in WoT applications, the MCSC framework utilises AES encryption, lightweight synchronization, and channel randomization. This study shows that achieving secure communication in WoT networks is possible without the major computational and energy costs linked to cryptographic techniques such as ECC. Applying AES encryption provides a powerful defence against frequent attacks such as MITM and eavesdropping and fits the needs of resource-limited settings like sensor systems and smart homes. Deploying the framework on Arduino Uno and nRF24N01 modules showed its realistic application. Evaluation demonstrated that MCSC guarantees stable communication characterised by minimal latency and a favourable rate of packet delivery together with a tolerable time for channel changes. This illustrates the feasibility of the framework in actual circumstances where energy efficiency and fast performance are crucial. Although MCSC provides an encouraging method, there are still some limits and opportunities for future exploration. Even though synchronisation is efficient, it can be improved more in settings where nodes move often and the network topology varies regularly. AES encryption guarantees effective protection, but it is worthwhile to consider quantum cryptographic strategies that may lower overhead. MCSC delivers an effective and affordable solution to safeguard the networks of the WoT. The upcoming work shall emphasise boosting scalability and enhancing synchronisation efficiency as well as looking into adaptive security solutions that change based on the distinct needs and challenges of different WoT applications. The proposed work establishes a base for innovative, secure communication systems essential for the safety and performance of future wireless connected devices.

**Author contributions** Prokash Barman is the lead author and wrote the main manuscript and text . All authors reviewed the manuscript and contributed in developing the main idea.

**Funding** The authors have not disclosed any funding.

**Data availability** No datasets were generated or analysed during the current study.

## Declarations

**Conflict of interest** The authors declare no competing interests.

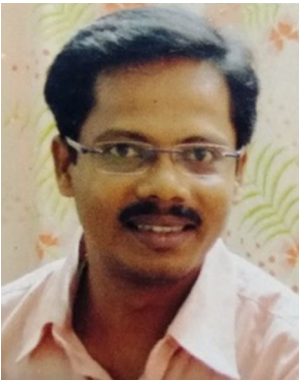

**Prokash Barman**, an experienced academic and technology expert, is the current Professor of Practice in the Department of Computational Sciences at Brainware University, India. He is a Ph.D. Research Fellow at the University of Calcutta and spends his time studying cybersecurity tactics and designing systems for the security of communication within IoT & IoMT. Over the past 18 years, he has acted as a Block Informatics Officer and supported digital technological growth in the West Bengal government, India. His experience involves IoT, Internet of Medical Things, communication security, cryptography, monitoring intrusions and designing hardware-software units. He has published several research papers in various SCI and Scopus-indexed journals and conferences. He is authoring three books on IoT, IoT Security and Power BI with BPB publication. His research has resulted three published Indian patents: Multi-Channel Secure Communication (MCSC) for Wireless of Things (WoT), a Lightweight Personal Body Scanner (PBS) and alert system for monitoring patients remotely and predicting illness and the Smart Ambulance Framework for Emergency Response with Embedded Monitoring System (SAFER-EMS). He regularly helps and guides others in his field and the impact of his research is strong in security, healthcare technology and new IoT solutions.

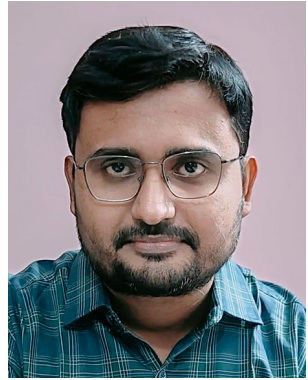

**Dr. Ratul Chowdhury** is an Assistant Professor in the Department of Computer Science and Engineering (AIML) at Netaji Subhash Engineering College under MAKAUT. He was awarded his Ph.D. in Computer Science and Engineering with a specialization in Network Security from the University of Calcutta in 2024. His research interests include Network Intrusion Detection Systems, Cloud Security, and IoT Security. Dr. Chowdhury has published over 10 research papers in various SCI and Scopus-indexed journals and conferences. With nearly a decade of teaching experience, he has served at several reputed institutions, including the Institute of Engineering and Management and the Future Institute of Engineering and Management in Kolkata.

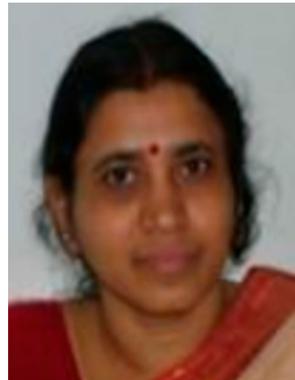

**Dr. Banani Saha** is a retired Professor of the Department of Computer Science and Engineering at the University of Calcutta. At the University of Calcutta, she studied and researched advanced computing and different disciplines for her Ph.D. (Tech) degree. Among her areas of research are as IoT, IoMT, Cryptography, IDS and Cybersecurity , as well as Data Mining, Data Warehousing, Image Processing, Data Fusion. Dr. Saha is concerned with integrating sophisticated security into both limited environmental resources and healthcare applications. Many postgraduate and doctoral students under her supervision today contribute to academia and industry, thanks to Dr. Saha's strong dedication to research and education. Her efforts have been shared in established journals and conferences, giving important contributions to computing and security research. Her leadership plays a major role in researching and developing even stronger secure and intelligent systems which helps shape the field of data and security technology.